\documentclass[aps,prc,preprint,superscriptaddress,showpacs]{revtex4}

\usepackage[dvips]{graphicx}
\usepackage{amsmath,amssymb,times}
\usepackage{ulem}

\def\lessa{\mathrel{\mathpalette\fun <}}
\def\ga{\mathrel{\mathpalette\fun >}}
\def\fun#1#2{\lower3.6pt\vbox{\baselineskip0pt\lineskip.9pt
\ialign{$\mathsurround=0pt#1\hfil##\hfil$\crcr#2\crcr\sim\crcr}}}

\newcommand{\beq}{\begin{equation}}
\newcommand{\eeq}{\end{equation}}
\newcommand{\bea}{\begin{eqnarray}}
\newcommand{\eea}{\end{eqnarray}}

\def\gsim{~\,\makebox(1,1){$\stackrel{>}{\widetilde{}}$}\,~}

\DeclareSymbolFont{boldletters}{OML}{cmm} {b}{it}
\DeclareSymbolFontAlphabet{\mathbit}{boldletters}
\DeclareMathSymbol{\alpha}{\mathalpha}{letters}{"0B}
\DeclareMathSymbol{\beta}{\mathalpha}{letters}{"0C}
\DeclareMathSymbol{\gamma}{\mathalpha}{letters}{"0D}
\DeclareMathSymbol{\delta}{\mathalpha}{letters}{"0E}
\DeclareMathSymbol{\epsilon}{\mathalpha}{letters}{"0F}
\DeclareMathSymbol{\zeta}{\mathalpha}{letters}{"10}
\DeclareMathSymbol{\eta}{\mathalpha}{letters}{"11}
\DeclareMathSymbol{\theta}{\mathalpha}{letters}{"12}
\DeclareMathSymbol{\iota}{\mathalpha}{letters}{"13}
\DeclareMathSymbol{\kappa}{\mathalpha}{letters}{"14}
\DeclareMathSymbol{\lambda}{\mathalpha}{letters}{"15}
\DeclareMathSymbol{\mu}{\mathalpha}{letters}{"16}
\DeclareMathSymbol{\nu}{\mathalpha}{letters}{"17}
\DeclareMathSymbol{\xi}{\mathalpha}{letters}{"18}
\DeclareMathSymbol{\pi}{\mathalpha}{letters}{"19}
\DeclareMathSymbol{\rho}{\mathalpha}{letters}{"1A}
\DeclareMathSymbol{\sigma}{\mathalpha}{letters}{"1B}
\DeclareMathSymbol{\tau}{\mathalpha}{letters}{"1C}
\DeclareMathSymbol{\upsilon}{\mathalpha}{letters}{"1D}
\DeclareMathSymbol{\phi}{\mathalpha}{letters}{"1E}
\DeclareMathSymbol{\chi}{\mathalpha}{letters}{"1F}
\DeclareMathSymbol{\psi}{\mathalpha}{letters}{"20}
\DeclareMathSymbol{\omega}{\mathalpha}{letters}{"21}
\DeclareMathSymbol{\varepsilon}{\mathalpha}{letters}{"22}
\DeclareMathSymbol{\vartheta}{\mathalpha}{letters}{"23}
\DeclareMathSymbol{\varpi}{\mathalpha}{letters}{"24}
\DeclareMathSymbol{\varrho}{\mathalpha}{letters}{"25}
\DeclareMathSymbol{\varsigma}{\mathalpha}{letters}{"26}
\DeclareMathSymbol{\varphi}{\mathalpha}{letters}{"27}
\DeclareMathSymbol{\Gamma}{\mathalpha}{letters}{"00}
\DeclareMathSymbol{\Delta}{\mathalpha}{letters}{"01}
\DeclareMathSymbol{\Theta}{\mathalpha}{letters}{"02}
\DeclareMathSymbol{\Lambda}{\mathalpha}{letters}{"03}
\DeclareMathSymbol{\Xi}{\mathalpha}{letters}{"04}
\DeclareMathSymbol{\Pi}{\mathalpha}{letters}{"05}
\DeclareMathSymbol{\Sigma}{\mathalpha}{letters}{"06}
\DeclareMathSymbol{\Upsilon}{\mathalpha}{letters}{"07}
\DeclareMathSymbol{\Phi}{\mathalpha}{letters}{"08}
\DeclareMathSymbol{\Psi}{\mathalpha}{letters}{"09}
\DeclareMathSymbol{\Omega}{\mathalpha}{letters}{"0A}

\begin{document}
\preprint{SAGA-HE-260-10}
\title{Investigation of meson masses for real and imaginary chemical potential 
\\
using the three-flavor PNJL model}

\author{Takeshi Matsumoto}
\email[]{t-matsumoto@phys.kyushu-u.ac.jp}
\affiliation{Department of Physics, Graduate School of Sciences, Kyushu University, Fukuoka 812-8581, Japan}

\author{Kouji Kashiwa}
\email[]{kashiwa@phys.kyushu-u.ac.jp}
\affiliation{Department of Physics, Graduate School of Sciences, Kyushu University, Fukuoka 812-8581, Japan}
             
\author{Hiroaki Kouno}
\email[]{kounoh@cc.saga-u.ac.jp}
\affiliation{Department of Physics, Saga University, Saga 840-8502, Japan}

\author{Kagayaki Oda}
\email[]{oda@phys.kyushu-u.ac.jp}
\affiliation{Department of Physics, Graduate School of Sciences, Kyushu University, Fukuoka 812-8581, Japan}

\author{Masanobu Yahiro}
\email[]{yahiro@phys.kyushu-u.ac.jp}
\affiliation{Department of Physics, Graduate School of Sciences, Kyushu University, Fukuoka 812-8581, Japan}

\date{\today}

\begin{abstract}
We investigate chemical-potential ($\mu$) and temperature ($T$) dependence of scalar and pseudo-scalar meson masses for both real and imaginary $\mu$, using the Polyakov-loop extended Nambu--Jona-Lasinio (PNJL) model with three-flavor quarks. 
A three-flavor phase diagram is drawn in $\mu^2$-$T$ plane where positive (negative) $\mu^2$ corresponds to positive (imaginary) $\mu$. A critical surface is plotted as a function of light- and strange-quark current mass and $\mu^2$. 
We show that $\mu$-dependence of the six-quark Kobayashi-Maskawa-'t Hooft (KMT) determinant interaction originated in $U_\mathrm{A}(1)$ anomaly can be determined from lattice QCD data on $\eta'$ meson mass around $\mu =0$ and $\mu = i \pi T/3$ with $T$ slightly above the critical temperature at $\mu=0$ where the chiral symmetry is restored at $\mu=0$ but broken at $\mu =i \pi T/3$, if it is measured in future. 
\end{abstract}

\maketitle

\section{Introduction}

In recent theoretical studies, novel scenarios for QCD phase structure at finite real chemical potential ($\mu_\mathrm{R}$) are suggested; for example, the quarkyonic phase~\cite{McLerran1,Hidaka,McLerran2,Miura}, the multi critical-endpoint generation~\cite{Kitazawa,Hatsuda,Zhang,Harada} and the Lifshitz-point induced by the inhomogeneous phase~\cite{Nickel}.  
Thus, {\it qualitative or speculative} investigation of QCD phase diagram is progressing well.

Nevertheless, {\it quantitative or more conclusive} understanding of QCD phase diagram is quite poor. 
The principal reason is the sign problem in the first-principle lattice QCD simulation at finite $\mu_\mathrm{R}$. 
Several methods such as the reweighting method~\cite{Fodor}, the
Taylor expansion method~\cite{Allton}, the analytic continuation from imaginary chemical potential $\mu_\mathrm{I}$ to $\mu_\mathrm{R}$~\cite{FP,Elia1,Elia2} and so on were proposed so far to circumvent the sign problem. 
However, they do not reach the $\mu_\mathrm{R}/T \gsim 1$ region yet. 
For this reason, effective models such as the Nambu--Jona-Lasinio (NJL) model were used so far to investigate qualitative properties of the phase structure at finite $\mu_\mathrm{R}$. 
The effective-model approach, however, has an ambiguity particularly in the interaction part; see Ref.~\cite{Kashiwa1} and references therein.

Thus, a new approach should be proposed for quantitative or more reliable investigation of QCD phase diagram at finite $\mu_\mathrm{R}$. 
As a possible answer, recently, we proposed the {\it imaginary chemical potential matching approach} 
(the $\mu_\mathrm{I}$-matching approach)~\cite{Kashiwa2,Sakai1}. 
In this approach, interactions of the effective model are determined from LQCD data at finite $\mu_\mathrm{I}$ where no sign problem comes out. After the determination, a phase structure at finite $\mu_\mathrm{R}$ is predicted with the effective model. 
The most important point in this approach is whether the model taken can reproduce the Roberge-Weiss (RW) periodicity and the RW transition at finite $\mu_\mathrm{I}$~\cite{RW}. 
In our previous works~\cite{Sakai1}, 
we showed that the Polyakov-loop extended Nambu--Jona-Lasinio (PNJL) model~\cite{Fukushima1} can do it, because the thermodynamical potential of the PNJL model is invariant under the extended ${\mathbb Z}_3$ transformation of 
\begin{align}
&e^{\pm i \theta} \to e^{\pm i \theta} e^{\pm i{2\pi k\over{3}}},~~~~  
 \Phi(\theta)  \to \Phi(\theta) e^{-i{2\pi k\over{3}}}, ~~~~
{\bar \Phi}(\theta) \to {\bar \Phi}(\theta) e^{i{2\pi k\over{3}}},
\label{eq:K2}
\end{align}
where $\theta=\mu_\mathrm{I}/T$. 
Here, $\Phi$ and ${\bar \Phi}$ denote the Polyakov-loop and its conjugate, respectively. This symmetry ensures the RW periodicity. 
Since, the PNJL model is designed to treat the confinement mechanism approximately in addition to the chiral symmetry breaking, we can investigate not only the chiral transition but also the deconfinement transition with the PNJL model. 
We also showed by using the PNJL model that the crossover deconfinement transition that takes place at finite $\theta$ becomes stronger as $\theta$ increases and eventually at $\theta =\pi/3$ it changes into the RW phase transition~\cite{Sakai1}.

The $U_\mathrm{A}(1)$ anomaly related to instantons can be taken into account in the NJL and PNJL models. In the three-flavor case, it is described by the effective six-quark Kobayashi-Maskawa-'t Hooft (KMT) determinant interaction~\cite{KM,'t Hooft}. 
The $U_\mathrm{A} (1)$ anomaly restoration at finite $T$ in the case of $\mu=0$ is investigated by the NJL model~\cite{Fukushima2} that can reproduce the lattice QCD data~\cite{Alles}.
At finite $\mu_\mathrm{R}$, $\mu$-dependence of the anomaly restoration strongly depends on that of coupling constant $G_D$ of the KMT determinant interaction~\cite{Chen}. 
However, $\mu$-dependence of $G_D$ is unclear, because lattice QCD data is not feasible at finite $\mu_\mathrm{R}$ and also theoretical understanding 
on $\mu$-dependence of the instanton density is not sufficient. 
Therefore, the phase structure in the three-flavor system is more ambiguous than in the two-flavor system.

In this paper, we investigate scalar and pseudo-scalar meson masses in both the $\mu_\mathrm{R}$ and $\mu_\mathrm{I}$ regions, using the three-flavor PNJL model. 
We show $\eta'$ meson mass is sensitive to $G_D$ particularly near 
$\theta=\pi/3$. This means that the $\theta$ dependence of $\eta'$ meson mass 
is a good quantity to determine $\mu$ dependence of $G_D$.
At the present stage, there is no reliable lattice QCD data particular on meson masses for the case of finite $\mu_\mathrm{I}$. 
Therefore, our investigation is limited to only a qualitative level.

\section{Three-flavor PNJL model}

Lagrangian density of the three-flavor PNJL model is
\begin{align}
 {\cal L}_{\rm PNJL}  
=& {\bar q}(i \gamma_\nu D^\nu - {\hat m_0} )q  
  + G_S \sum_{a=0}^{8} 
    [({\bar q} \lambda_a q )^2 +({\bar q }i\gamma_5 \lambda_a q )^2] 
\nonumber\\
 &- G_D \Bigl[\det_{ij} {\bar q}_i (1+\gamma_5) q_j 
           +\det_{ij} {\bar q}_i (1-\gamma_5) q_j \Bigr]
  -{\cal U}(\Phi [A],{\bar \Phi} [A],T) , 
\label{L}
\end{align} 
where $D^\nu=\partial^\nu + iA^\nu=\partial^\nu +i\delta^{\nu}_{0}gA^0_a{\lambda_a / 2}$ with the gauge coupling $g$ and the Gell-Mann matrices $\lambda_a$. 
Three-flavor quark fields $q=(q_u,q_d,q_s)$ have current quark masses 
${\hat m_0}={\rm diag}(m_u,m_d,m_s)$. The Polyakov potential $\cal{U}$ is defined later in \eqref{eq:E13} and \eqref{eq:E14}. 
In the interaction part, $G_S$ and $G_D$ denote coupling constants 
of the scalar-type four-quark and the KMT determinant interaction, 
respectively. 
The determinant $\displaystyle \det_{ij}$ runs in the flavor space and then the KMT determinant interaction breaks the $U_\mathrm{A} (1)$ symmetry explicitly.

In the PNJL model, the gauge field $A_\mu$ is treated as a homogeneous and static background field. 
The Polyakov-loop $\Phi$ and its conjugate ${\bar \Phi}$ are given by
\begin{align}
\Phi &= {1\over{3}}{\rm tr}_{\rm c}(L),
~~~~~\bar{\Phi} ={1\over{3}}{\rm tr}_{\rm c}({\bar L})
\label{Polyakov}
\end{align}
where $L  = \exp(i A_4/T)$ with $A_4=iA_0$ in Euclidean space. 
In the Polyakov-gauge, $A_4$ is diagonal in the color space.

We make the mean field approximation (MFA) to the quark-quark interactions in \eqref{L} in the following way. 
In \eqref{L}, the operator product ${\bar q}_iq_j$ is first divided into $ {\bar q}_i q_j = \sigma_{ij}  + ({\bar q}_i q_j)'$ with the mean field $\sigma_{ij} \equiv \langle {\bar q}_i q_j \rangle$ and the fluctuation $({\bar q}_i q_j)'$ where $i,j=u,d,s$. 
Ignoring higher-order terms of $({\bar q}_i q_j)'$ in the rewritten Lagrangian and re-substituting 
$ ({\bar q}_i q_j)' = {\bar q}_i q_j -\sigma_{ij} $ into the approximated Lagrangian, one can obtain a linearized Lagrangian based on MFA: 
\begin{align}
 {\cal L}_{\rm PNJL}^{\rm MFA}  
=& {\bar q}_i(i \gamma_\nu \partial^\nu+i\gamma_0A_4 -M_{ii})q_i
 -\Bigl(
   \sum_{i=u,d,s}  2 G_S \sigma_{ii}^2 
 - 4 G_D \sigma_{uu}\sigma_{dd}\sigma_{ss} \Bigr)
\nonumber\\
 & -{\cal U}(\Phi [A],{\bar \Phi} [A],T),
\label{linear-L}
\end{align} 
where the dynamical quark mass $M_{ii}$ is defined by 
$M_{ii}=m_{i}-4G_S\sigma_{ii}+  2G_D \sigma_{jj} \sigma_{kk}$ with $i \neq j \neq k$. 

In this study, we impose the isospin symmetry for ${u}$-${d}$ sector and then we use $m_{l}=m_{u}=m_{d}$. 
The thermodynamical potential becomes
\begin{align}
\Omega_\mathrm{PNJL} 
&= -2 \sum_{f=u,d,s} \int \frac{d^3 p}{(2\pi)^3}
   \Bigl[ N_\mathrm{c} E_{p,f} \nonumber\\
&        + \frac{1}{\beta}
           \ln~ [1 + 3(\Phi+{\bar \Phi} e^{-\beta (E_{p,f}-\mu_{f})}) e^{-\beta (E_{p,f}-\mu_f)}+ e^{-3\beta (E_{p,f}-\mu_f)}] \notag\\
&        + \frac{1}{\beta} 
           \ln~ [1 + 3({\bar \Phi}+\Phi e^{-\beta (E_{p,f}+\mu_{f})}) e^{-\beta (E_{p,f}+\mu_f)}+ e^{-3\beta (E_{p,f}+\mu_f)}]
	      \Bigl]
	      \nonumber\\
&+ \Bigl( \sum_{i=u,d,s}  2 G_S \sigma_{ii}^2 
 - 4 G_D \sigma_{uu}\sigma_{dd}\sigma_{ss} \Bigr)+{\cal U}(\Phi [A],{\bar \Phi} [A],T).
\label{PNJL-Omega}
\end{align}
We take the three-dimensional momentum cutoff,
\begin{equation}
\int \frac{d^3 p}{(2 \pi)^3}\to 
{1\over{2\pi^2}} \int_0^\Lambda dp p^2,
\end{equation}
because this model is non-renormalizable. 
Hence, the present model has five parameters $G_S$, $G_D$, $m_l$, $m_s$ and $\Lambda$.
We use the parameter set determined in Ref.~\cite{Klevansky}; these are fitted to the empirical values of $\pi$ meson mass and its decay constant, $K$ meson mass and its decay constant and $\eta'$ meson mass. We also use ${\cal U}$ of Ref.~\cite{Ratti} fitted to LQCD data in the pure gauge limit at finite $T$~\cite{Boyd,Kaczmarek}: 
\begin{align}
&{{\cal U}\over{T^4}} =  -\frac{b_2(T)}{2} {\bar \Phi}\Phi
              -\frac{b_3}{6}({\bar \Phi}^3+ \Phi^3)
              +\frac{b_4}{4}({\bar \Phi}\Phi)^2, \label{eq:E13} \\
&b_2(T)   = a_0 + a_1\Bigl(\frac{T_0}{T}\Bigr)
                 + a_2\Bigl(\frac{T_0}{T}\Bigr)^2
                 + a_3\Bigl(\frac{T_0}{T}\Bigr)^3.
\label{eq:E14} 
\end{align}
In this study, we take the original value $T_0=270$ MeV.

\section{Meson mass formalism}

First, Lagrangian density \eqref{linear-L} is rewritten by MFA into 
\begin{align}
{\cal L} &= {\bar q}_i (i \gamma_\nu D^\nu -{\hat m_0}) q_j
+\sum_{a=0}^8 [G^-_a ({\bar q} \lambda_a q)^2
              +G^+_a ({\bar q} i \gamma_5 \lambda_a q)^2 ] \nonumber\\
&+\sum_{a,b=0,3,8} [G^-_{ab} ({\bar q} \lambda_a q) ({\bar q} \lambda_b q)
                  +G^+_{ab} ({\bar q} i \gamma_5 \lambda_a q)
                         ({\bar q} i \gamma_5 \lambda_b q) ] 
\end{align}
with 
\begin{align}
\left\{
\begin{array}{cc}
  G_0^\pm = G_S \mp \dfrac{1}{3} G_D (\sigma_{uu} + \sigma_{dd} + \sigma_{ss}), 
& G_1^\pm = G_2^\pm = G_3^\pm
          = G_S \pm \dfrac{1}{2} G_D \sigma_{ss}, 
\vspace{.5em}\\
  G_4^\pm = G_5^\pm = G_S \pm \dfrac{1}{2} G_D \sigma_{dd}, 
& G_6^\pm = G_7^\pm = G_S \pm \dfrac{1}{2} G_D \sigma_{uu}, 
\vspace{.5em}\\
  G_8^\pm = G_S \pm \dfrac{1}{6} G_D ( 2\sigma_{uu} + 2\sigma_{dd} - \sigma_{ss}),  
& G_{30}^\pm = G_{03}^\pm = \mp \dfrac{1}{2\sqrt{6}} G_D (\sigma_{uu}-\sigma_{dd}), 
\vspace{.5em}\\
  G_{08}^\pm = G_{80}^\pm 
  = \pm \dfrac{\sqrt{2}}{12} G_D ( \sigma_{uu} + \sigma_{dd} - 2\sigma_{ss}),  
& G_{38}^\pm = G_{83}^\pm = \pm \dfrac{1}{2\sqrt{3}} G_D(\sigma_{uu}-\sigma_{dd}).
\vspace{.5em}
\end{array}
\right.
\label{EC-Meson}
\end{align}
In the present case that the isospin symmetry is imposed in the $u$-$d$ sector, we have $G_{30}^\pm = G_{03}^\pm = G_{38}^\pm = G_{83}^\pm=0$.

Taking the same procedure as in the two-flavor case~\cite{Kashiwa2}, we can obtain dynamical meson masses from poles of the effective propagator~\cite{Klevansky}: 
\begin{align}
\frac{2iG_{i}}
{1 - 2G_{i} \Pi(q_0)} 
\to \Bigl( 1 - 2G_{i} \Pi (q_0) \Bigr)
\Bigl|_{q_0=M_{\xi}} = 0
\label{mmf}
\end{align}
where $\Pi$ denotes the Polarization function. 
The subscript $i$ stands for meson $\xi$ in a state $i$. 
The polarization function between states $i$ and $j$ is represented by 
\begin{align}
\Pi_{ij} (q)
&= -i\int \frac{d^4p}{(2\pi)^4} {\rm Tr} 
   \Bigl[ \Gamma(T_i) S_F(p) \Gamma(T_j) S_F(p-q) \Bigr],
\label{eq:pf}
\end{align}
where $S_{F}$ denotes the quark propagator and  the vertex function $\Gamma$ is ${\bf 1}$ for the scalar meson and $i\gamma_5$ for the pseudo-scalar meson. 
The matrices $T_i$ in flavor space depend on meson considered; for example, $T_i=T_j=\lambda_3$ for $\pi$ and $a_0$ mesons and $T_i=(\lambda_6+i\lambda_7)/\sqrt{2},T_j=(\lambda_6+i\lambda_7)/\sqrt{2}$ 
for $K$ and $\kappa$ mesons.
When $T$ and $\mu$ are finite, the corresponding equations are obtained by the replacement 
\begin{align}
&p_0 \to i \omega_n + \mu -iA_4= i \pi T(2n+1)  +\mu -iA_4, 
\nonumber\\
&\int \frac{d^4p}{(2 \pi)^4} 
\to iT\sum_n \int \frac{d^3p}{(2 \pi)^3}. 
\end{align}
As for $\eta$ and $\eta'$ mesons, the effective coupling constant 
$G^{\pm}$ and the polarization function $\Pi$ are $2\times2$ matrices in flavor space, since the isospin symmetry is imposed for the $u$-$d$ sector: 
\begin{eqnarray}
 G^{\pm} =
\left[
\begin{array}{cc}
 G^{\pm}_{00} & G^{\pm}_{08} \\
 G^{\pm}_{80} & G^{\pm}_{88} \\
\end{array}
\right],~~~~ 
 \Pi =
\left[
\begin{array}{cc}
 \Pi_{00}&\Pi_{08} \\
 \Pi_{80}&\Pi_{88} \\
\end{array}
\right],
\end{eqnarray}
where
\begin{align}
 \Pi_{00}=\frac{1}{3}\left[2\Pi^{ll}+\Pi^{ss}\right],~~~
 \Pi_{88}=\frac{1}{3}\left[\Pi^{ll}+2\Pi^{ss}\right],~~~
 \Pi_{08}=\Pi_{80} = \frac{\sqrt{2}}{6}\left[\Pi^{ll}-\Pi^{ss}\right] .
\end{align}
Here, the polarization function $\Pi^{ff}$ for each flavor $f$ is defined by 
\begin{align}
\Pi^{ff} (q)
&= -2i\int \frac{d^4p}{(2\pi)^4} {\rm Tr}_{C,D} 
   \Bigl[ \Gamma S_F^f(p) \Gamma S_F^f(p-q) \Bigr]
\label{eq:pff}
\end{align}
with $S_F^f(q)$ the propagator of quark with flavor $f$, where 
in ${\rm Tr}_{C,D}$ the trace is taken for color and Dirac indices.
Therefore, $\eta$ and $\eta'$ meson masses satisfy 
\begin{align}
\det (1 - 2G^{+} \Pi) = 0.
\label{eta-mass}
\end{align}
This equation has two solutions; the lower corresponds to $\eta$ meson mass, while the higher does to $\eta'$ meson mass. 
Masses of $\sigma$ and $f_0$ meson are obtained by replacing $G^{+}$ by $G^-$ in \eqref{eta-mass} and setting $\Gamma=1$ in (\ref{eq:pff}).

\section{Numerical results}

\begin{figure}[htbp]
\begin{center}
 \includegraphics[width=0.5\textwidth]{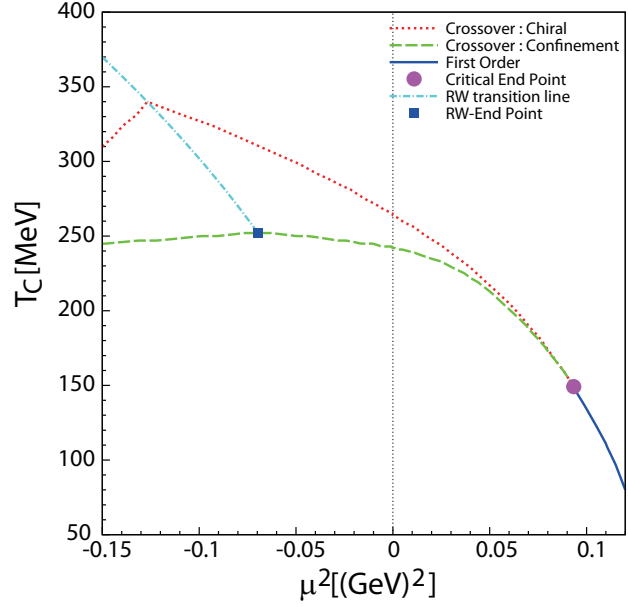}
\end{center}
\caption{(Color online) 
Phase diagram in $\mu^2-T$ plane for the three-flavor case. 
{
The solid (dotted) line denotes a first-order (crossover) chiral transition, while the dot-dashed (dashed) line does a first-order RW (crossover deconfinement) transition.  
The closed circle (square) stands for an endpoint of the first-order chiral (RW) phase transition. 
}
}
\label{Fig-PD}
\end{figure}

Figure~\ref{Fig-PD} shows the phase diagram in $\mu^2-T$ plane for the three-flavor case.
The positive (negative) $\mu^2$ half-plane means real (imaginary) $\mu$.  
In the present parameter set, a critical endpoint arises in the positive 
$\mu^2$ half-plane, while a RW endpoint does at $\mu^2=-(\pi T/3)^2$ with $T \approx 250$~MeV in the negative $\mu^2$ half-plane. 
At the RW endpoint, the phase transition is second order. 
In the two-flavor case, as shown in \cite{Sakai:2009vb}, the order of RW phase transition at RW endpoint depends on the Polyakov potential ${\cal U}$ taken; it is second order for ${\cal U}$ of Ref.~\cite{Fukushima1}, but first order for ${\cal U}$ of Ref.~\cite{Rossner}. 
And the latter gives a result more consistent with lattice QCD data at finite $\mu_{\rm I}$ than the former. 
This sort of analysis is quite important also for the three-flavor case in order to determine the form of ${\cal U}$, if precise lattice QCD data on the RW endpoint become available in future.

\begin{figure}[htbp]
\begin{center}
 \includegraphics[width=0.5\textwidth]{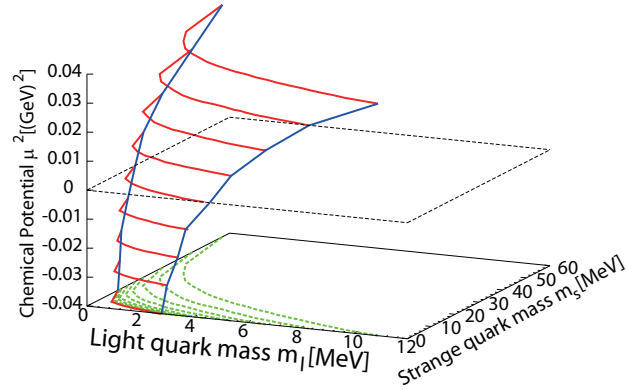}
\end{center}
\caption{(Color online) Critical surface 
as a function of $m_l$, $m_s$ and $\mu^2$ 
in the $\mu_\mathrm{R}$ and $\mu_\mathrm{I}$ regions.} 
\label{Fig-CF}
\end{figure}

The critical endpoint (closed circle) in Fig.~\ref{Fig-PD} is a function of $m_l$, $m_s$ and $\mu^2$. This is described as a surface in the $m_l$-$m_s$-$\mu^2$ space. The surface, usually called the critical surface, is plotted in Fig.~\ref{Fig-CF}. 
In the present model, the critical surface has a positive curvature and then a critical endpoint arises somewhere in the $\mu^2$-$T$ plane, as shown in Fig.~\ref{Fig-PD}, when $m_l$ and $m_s$ are taken to be physical values. 
This result may change, if the coupling constant $G_D$ of 
KMT determinant interaction depends on $\mu$~\cite{Chen}. 
However, the $\mu$-dependent $G_D$ of Ref.~\cite{Chen} can not be applicable to the imaginary $\mu$ region, since it breaks the extended ${\mathbb Z}_3$ symmetry, i.e. the RW periodicity. 
Thus, at the present stage, we have no way of determining $\mu$ dependence of $G_D$. We then do not consider any $\mu$-dependent $G_D$ in this paper.

\begin{figure}[htbp]
\begin{center}
 \includegraphics[width=0.42\textwidth]{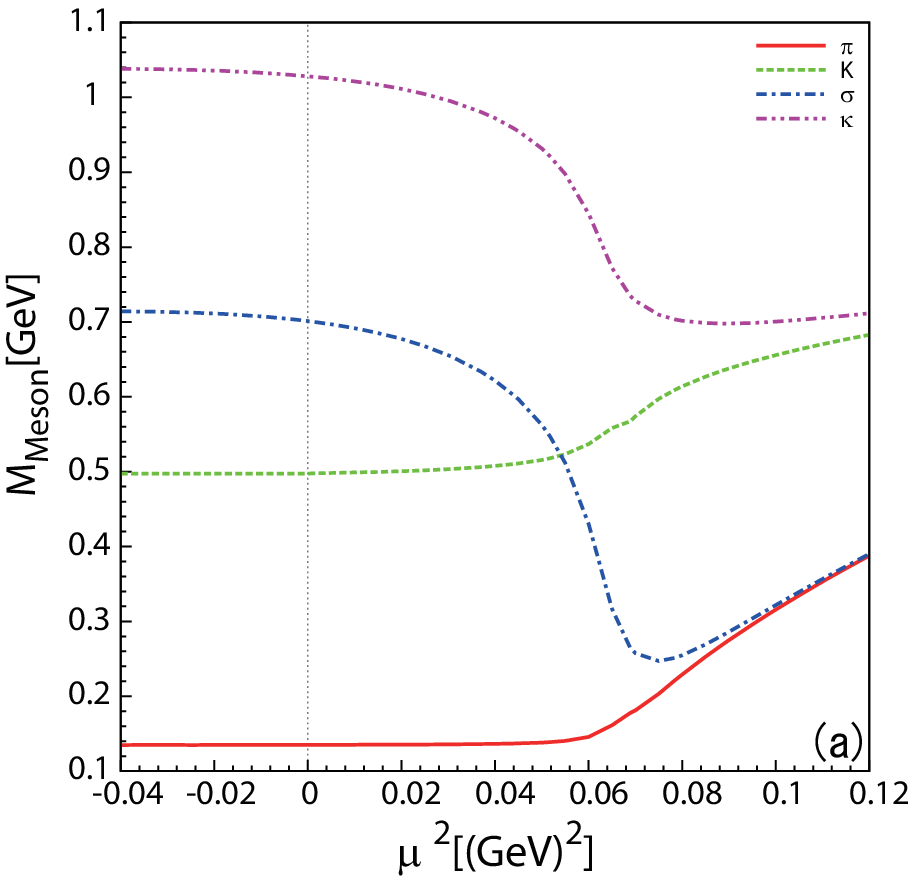}
 \includegraphics[width=0.4\textwidth]{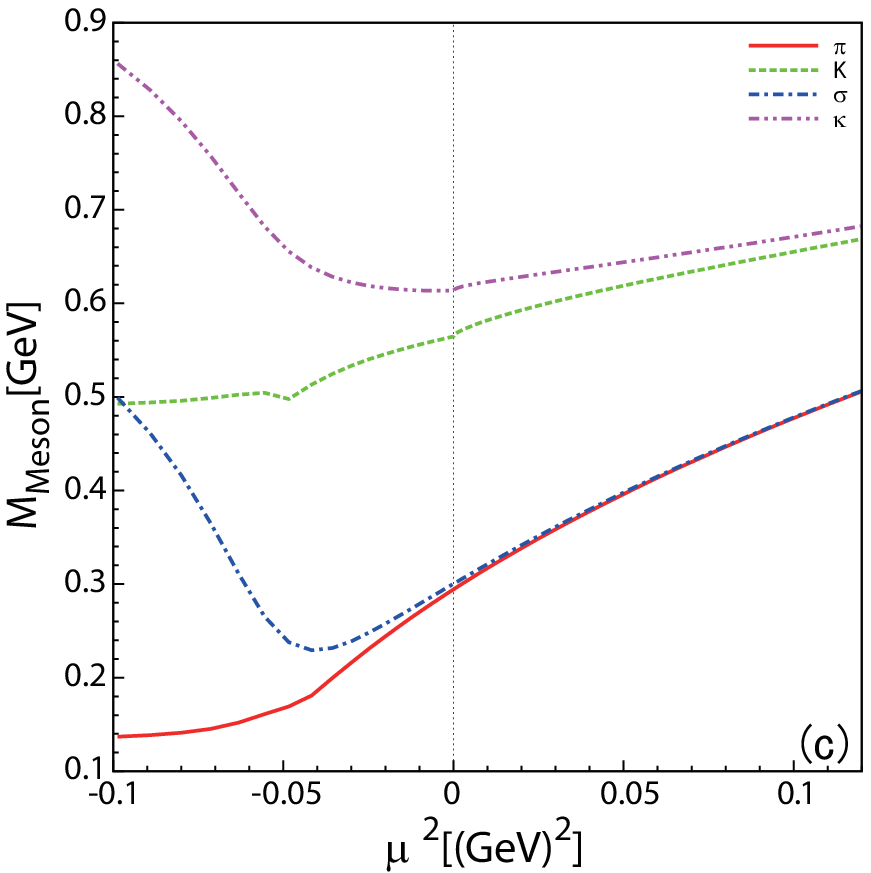}
 \includegraphics[width=0.42\textwidth]{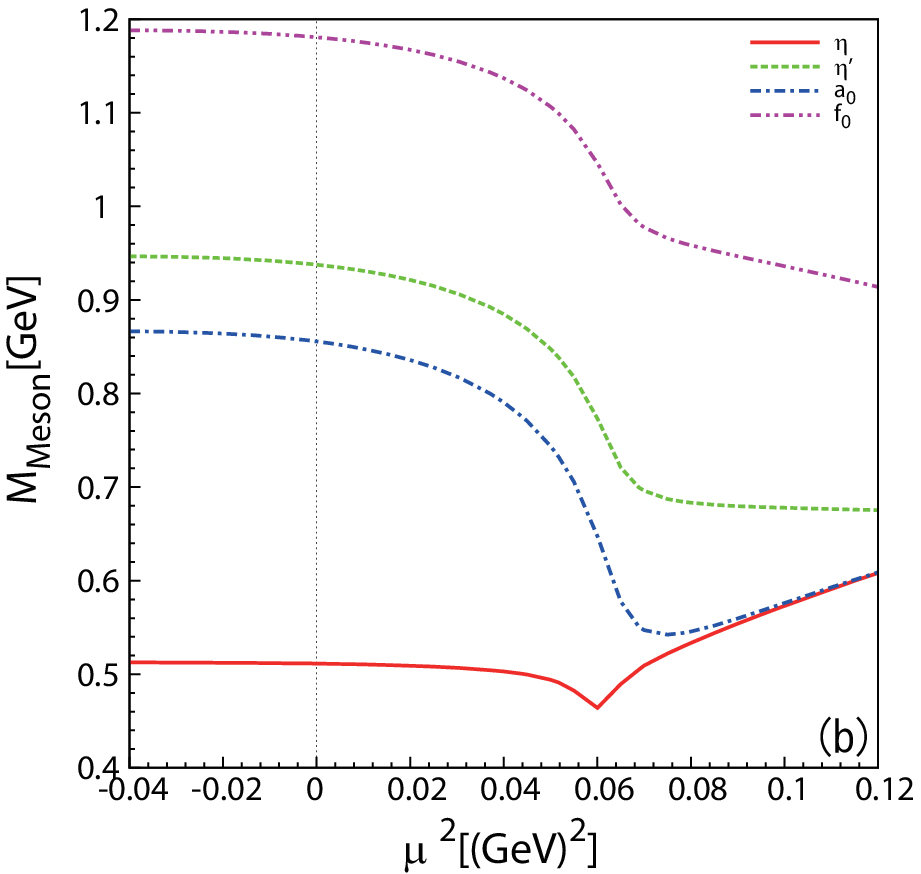}
 \includegraphics[width=0.4\textwidth]{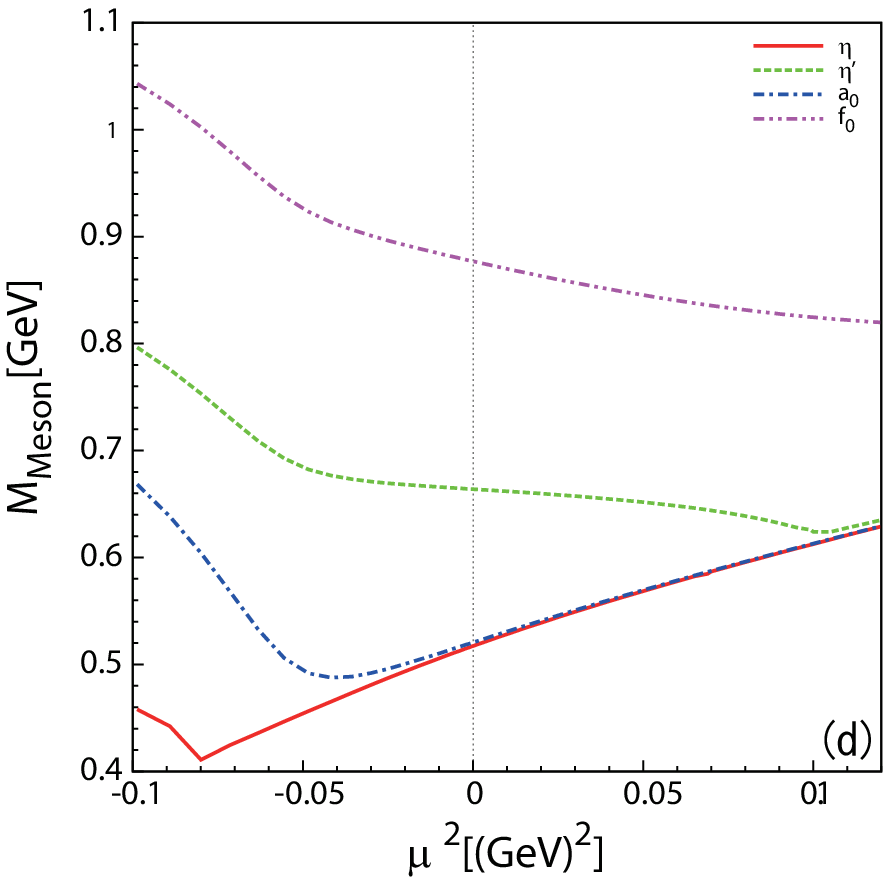}
\end{center}
\caption{ (Color online) 
$\mu^2$-dependence of meson mass for 
(a) $\pi$, $\sigma$, $K$ and $\kappa$ at $T=200$ MeV,
(b) $\eta$, $\eta'$, $a_0$ and $f_0$ at $T=200$ MeV, 
(c) $\pi$, $\sigma$, $K$ and $\kappa$ at $T=300$ MeV,
and
(d) $\eta$, $\eta'$, $a_0$ and $f_0$ at $T=300$ MeV. 
}
\label{Fig-mu2-Tdep}
\end{figure}

\begin{figure}[htbp]
\begin{center}
 \includegraphics[width=0.4\textwidth]{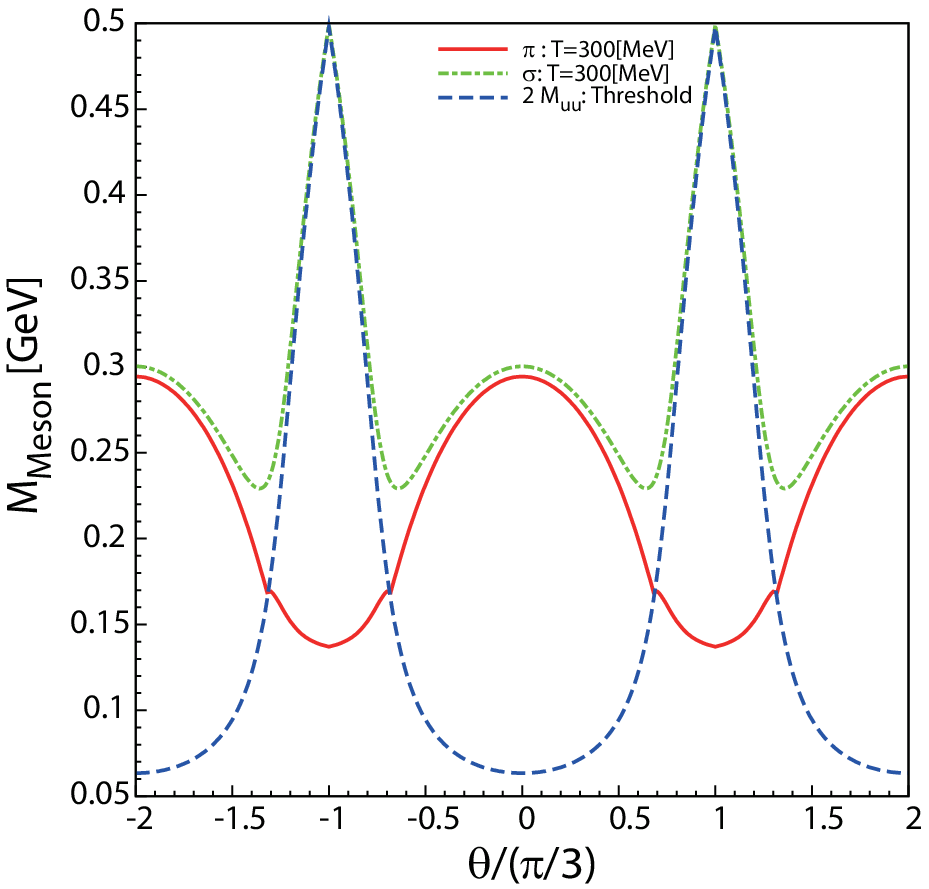}
 \includegraphics[width=0.4\textwidth]{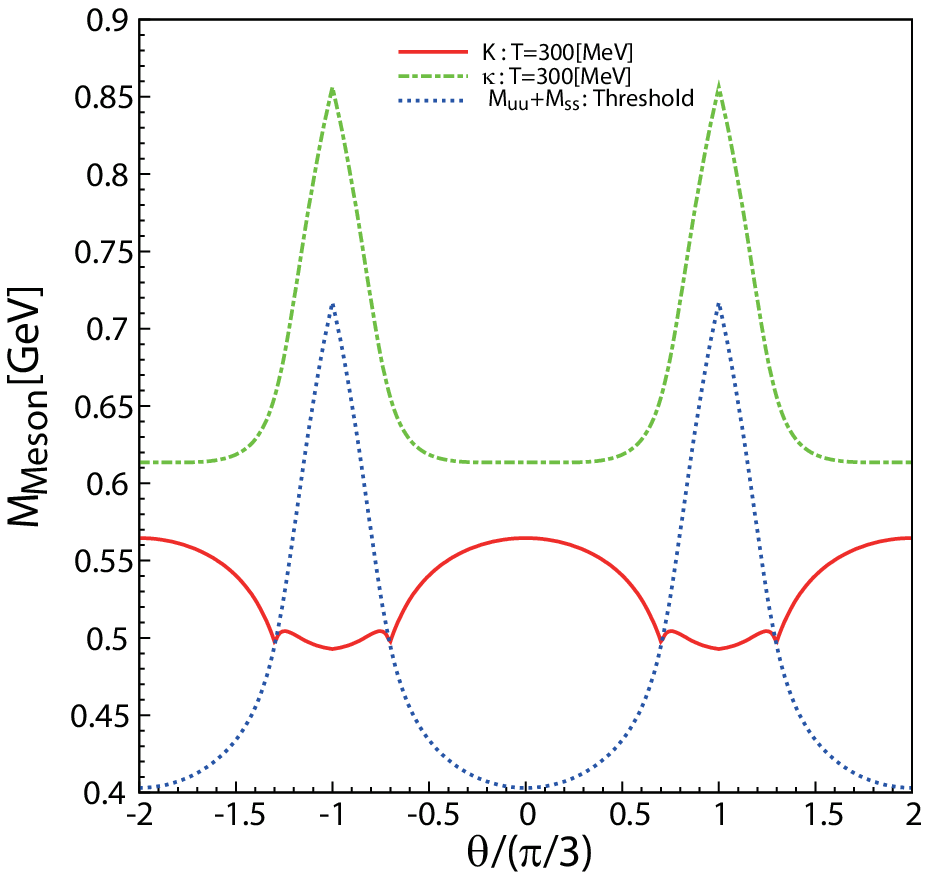}
\end{center}
\caption{(Color online) 
Meson mass at $T=300$ MeV as a function of $\theta/(\pi/3)$. 
The left panel corresponds to $\pi$ and $\sigma$, while the right panel does to $K$ and $\kappa$. 
}
\label{Fig-pskk-T300}
\end{figure}

In Fig.~\ref{Fig-mu2-Tdep}, we investigate $\mu^2$-dependence of meson mass for mesons ($\pi$, $K$, $\eta$, $\eta'$, $\sigma$, $\kappa$, $a_0$, $f_0$), using the PNJL model in which $G_D$ is a constant $G_D=G_D(0)=-12.36\Lambda^{-5}$ where $G_D(0)$ is determined at $T=\mu=0$. 
The left and the right panels correspond to $T=200$ and $300$ MeV, respectively. Furthermore, Figure~\ref{Fig-pskk-T300} shows $\theta$ dependence of meson mass for $T=300$ MeV; the left (right) panel corresponds to $\pi$ and $\sigma$ ($K$ and $\kappa$).  
Obviously, these meson masses are $\theta$-even and have the RW periodicity. 
Pion mass has dips at $\mu^2 \approx 0.06$~$({\rm GeV})^2$ in panel (b), at $\mu^2 \approx -0.08$~$({\rm GeV})^2$ in panel (d) of Fig.~\ref{Fig-mu2-Tdep} and at 
$\theta \approx \pm0.7$ and $\pm1.3$~MeV of Fig.~\ref{Fig-pskk-T300}. 
These are threshold effects due to $\pi \rightarrow {\rm quark} + {\rm antiquark}$.

A mass difference between the chiral partners $\pi$ and $\sigma$ is a good indicator of the chiral symmetry; the symmetry is restored (broken) when the difference is small (large). 
As for $T=300$~MeV, as shown in panel (c) of Fig.~\ref{Fig-mu2-Tdep} and the left panel of Fig.~\ref{Fig-pskk-T300}, the chiral symmetry is restored at $\mu^2 \ga -0.02$~(GeV)$^2$, but broken at $\mu^2 \lessa -0.02$~(GeV)$^2$ ($\pi/6 \lessa \theta \le \pi/3$). 
As for $T=200$~MeV, as shown in panel (a) of Fig.~\ref{Fig-mu2-Tdep}, the symmetry is restored at $\mu^2 \ga 0.1$~(GeV)$^2$, but broken at $\mu^2 \lessa 0.1$~(GeV)$^2$. 
As an interesting result in Fig.~\ref{Fig-mu2-Tdep}, panels (c) and (d) almost agree with panels (a) and (b), respectively, if in panels (c) and (d) the $\mu^2$ scale is shifted to the left by about $0.1$. 
Thus, shifting the $\mu^2$ scale to the left corresponds to looking at meson mass at lower $T$. If meson mass is measured in future by lattice QCD in the negative $\mu^2$ region for some temperature $T_{\rm Latt}$, the behavior qualitatively agrees with that in the positive $\mu^2$ region for temperature lower than $T_{\rm Latt}$. Therefore, we can predict qualitative behavior of meson mass in the positive $\mu^2$ region from lattice data on meson mass in the negative $\mu^2$ region.

\begin{figure}[htbp]
\begin{center}
 \includegraphics[width=0.4\textwidth]{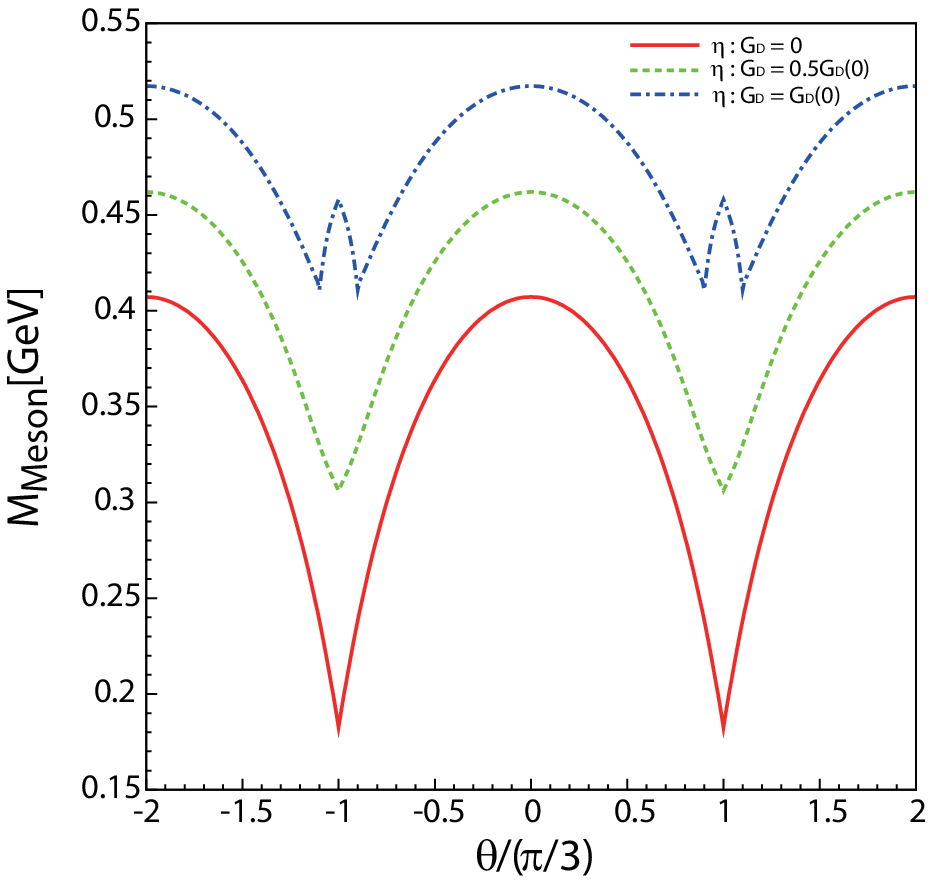}
 \includegraphics[width=0.4\textwidth]{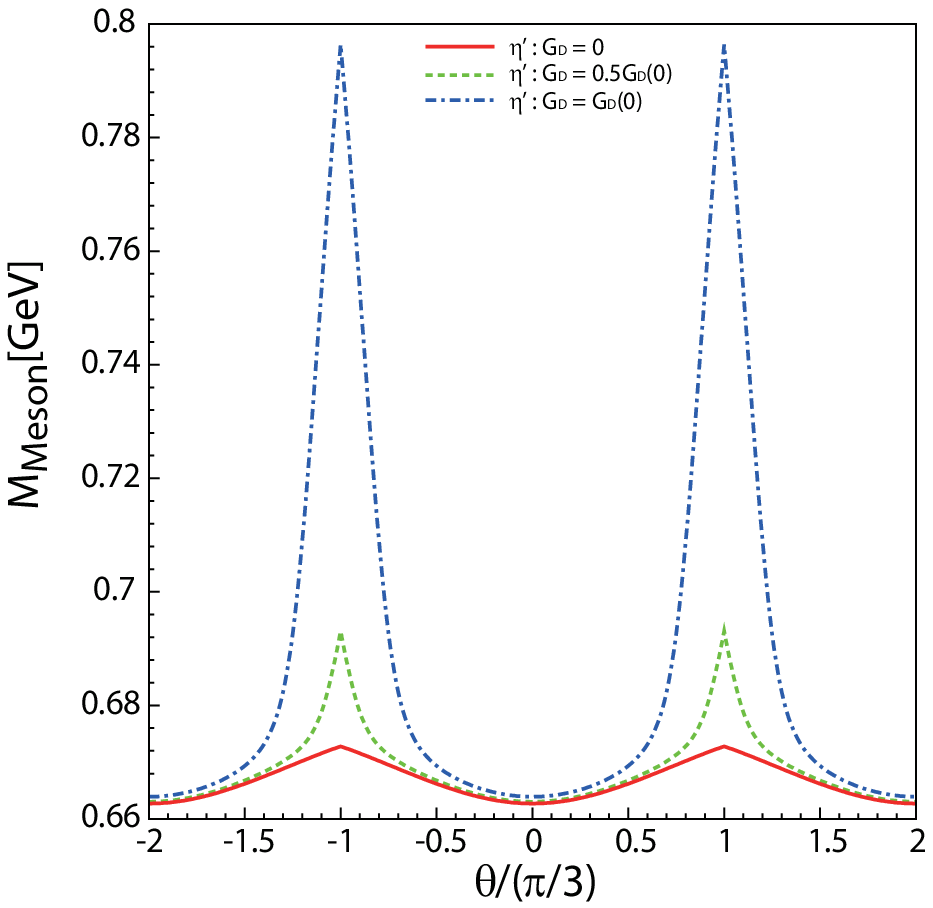}
\end{center}
\caption{(Color online)
$\theta$-dependence of meson mass at $T=300$ MeV for 
$\eta$ in left panel and $\eta'$ in right panel. 
Three values of $G_D=0$, $G_D(0)/2$ and $G_D(0)$ are taken. 
}
\label{Fig-hhp-T300-Kdep}
\end{figure}

Next, it is investigated how the KMT determinant interaction affects meson mass at imaginary $\mu$ 
by changing the value of $G_D$ from the original one $G_D(0)$. 
\begin{figure}[htbp]
\begin{center}
 \includegraphics[width=0.4\textwidth]{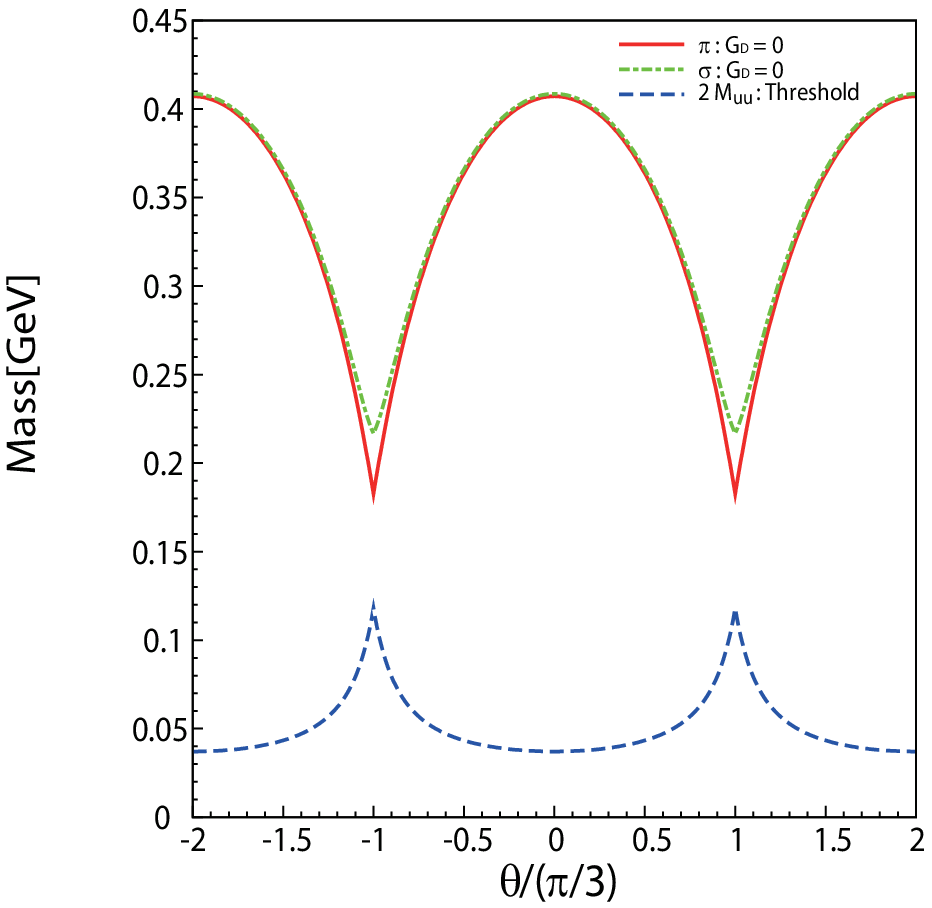}
 \includegraphics[width=0.4\textwidth]{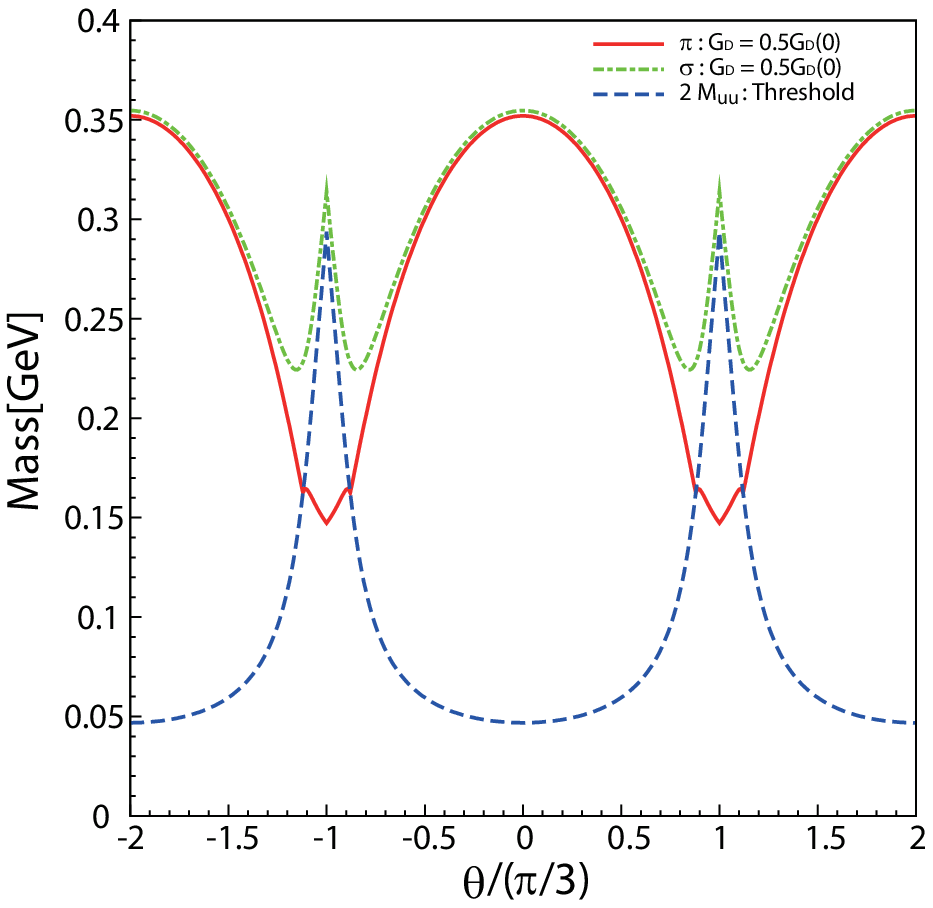}
\end{center}
\caption{ (Color online)
$\theta$-dependence of $\sigma$ and $\pi$ meson masses at $T=300$ MeV for 
$G_D=0$ in the left panel and for $G_D=G_D(0)/2$ in the right panel. 
}
\label{Fig-ps-T300-Kdep}
\end{figure}
As mentioned above, theoretically, 
$G_D$ is allowed to have $\mu$ dependence. 
However, since the actual form is unknown, we simply change the value of $G_D$ in the present analysis. 
Figure~\ref{Fig-ps-T300-Kdep} shows $\pi$ and $\sigma$ meson masses at $T=300$ MeV with the KMT determinant interaction in which 
$G_D=0$ for left panel and $G_D=G_D(0)/2$ for right panel. 
As $G_D$ increases, the $\pi$ meson mass is reduced with almost keeping the $\theta$ dependence. 
The $\sigma$ meson mass is also reduced, but the $\theta$ dependence is changed a lot around $\theta = \pi/3$. 
The left (right) panel of Fig.~\ref{Fig-hhp-T300-Kdep} shows $\eta$ ($\eta'$) meson mass at $T=300$ MeV in three cases of $G_D=0$, $G_D(0)/2$ and $G_D(0)$. 
The $\eta$ meson mass has a similar property to the $\sigma$ meson mass. 
Most interesting and important property is $\theta$ dependence of $\eta'$ meson mass. 
The $\eta'$ meson mass has a weak $G_D$ dependence at $\theta=0$, but the $G_D$ dependence becomes strong around $\theta = \pi/3$. Thus, 
$\theta$ dependence of $G_D$ that is allowed theoretically can be determined from that of $\eta'$ meson mass, if it is measured in future by lattice QCD.

The sensitivity of $\eta'$ meson mass to $G_D$ around $\theta=\pi/3$ can be understood in the following. 
The KMT determinant interaction affects the meson mass only through the dynamical quark mass $M_{ii}=m_{i}-4G_S\sigma_{ii}+  2G_D \sigma_{jj} \sigma_{kk}$ with a $G_D$ dependent term of form $G_D\sigma_{ii} \sigma_{jj}$. 
The term $G_D \sigma_{ii} \sigma_{jj}$ is strongly suppressed when the chiral symmetry is restored, even if $G_D$ is large. 
Hence, the $U_\mathrm{A} (1)$ anomaly affects the meson mass only when the chiral symmetry is broken. 
As shown in Fig.~\ref{Fig-PD}, the critical temperature of the chiral transition goes up as $\theta$ increases from 0 to $\pi/3$.
Thus, $\eta'$ meson mass is most sensitive to $G_D$ at $\theta = \pi/3$ when $\theta$ increases from 0 to $\pi/3$ with temperature fixed, because $\sigma_{ii}$ and $\sigma_{jj}$ are largest there.

In this study, $G_D$ is assumed to be constant. 
If $\eta'$ meson mass is measured by three-flavor lattice QCD in future, there is a possibility that the PNJL model with constant $G_D$ can not reproduce lattice QCD data. If so, the deviation can determine $\theta$-dependence of $G_D$ and hence 
$\mu_\mathrm{R}$-dependence of $G_D$.

\section{Summary}

Using the three-flavor PNJL model, we have analyzed $\mu$-dependence of scalar and pseudo-scalar meson masses in both the real and the imaginary $\mu$ region. 
In the imaginary $\mu$ region, the meson masses are even functions 
of $\theta$ with the RW periodicity. 
As an interesting result, $\mu^2$ dependence of meson mass in the negative 
$\mu^2$ region at some temperature is close to that in the positive $\mu^2$ region at temperatures lower than the temperature. 
If meson mass is measured in future by lattice QCD in the negative $\mu^2$ region for some temperature $T_{\rm Latt}$, the behavior qualitatively agrees with that in the positive $\mu^2$ region for temperatures lower than $T_{\rm Latt}$. 
Therefore, we can predict qualitative behavior of meson mass in the positive $\mu^2$ region from lattice data on meson mass in the negative $\mu^2$ region.

The $U_\mathrm{A}(1)$ anomaly (the KMT determinant interaction) affects meson masses through the term $G_D \sigma_{jj} \sigma_{kk}$ in the dynamical quark mass $M_{ii}$. 
Particularly, the effect is remarkable for $\eta'$ meson. 
For temperatures slightly above the critical temperature at $\theta=0$, the chiral condensate increases a lot as $\theta$ increases from 0 to $\pi/3$, so that the effect has a strong $\theta$ dependence. 
We then recommend that meson masses, particularly $\eta'$ meson mass, be measured by lattice QCD for $\theta = 0$ and $\pi /3$ at such higher temperatures. 
Using the lattice QCD data, we can determine $T$ and $\mu$ dependences of coupling constant $G_D$ of the KMT determinant interaction and hence can predict the three-flavor phase diagram with higher reliability by the PNJL model.

\noindent
\begin{acknowledgments}
The authors thank Dr. H. Kohyama for useful discussion.
K.K. is supported by the Japan Society for the Promotion of Science for Young Scientists.
\end{acknowledgments}


\end{document}